# Questions & Answers
# for TEI Newcomers
(september 20, 2008)

**Please send your text to:**

redaktion@computerphilologie.de
**malte.rehbein@nuigalway.ie**

## Abstract

This paper provides an introduction to the Text Encoding Initiative (TEI), focused at bringing in newcomers who have to deal with a digital document project and are looking at the capacity that the TEI environment may have to fulfil his needs. To this end, we avoid a strictly technical presentation of the TEI and concentrate on the actual issues that such projects face, with parallel made on the situation within two institutions. While a quick walkthrough the TEI technical framework is provided, the papers ends up by showing the essential role of the community in the actual technical contributions that are being brought to the TEI.

## Introduction

Most scholars in the humanities who have been in the situation of managing a textual source in digital format are aware of the existence of the TEI (Text Encoding Initiative, www.tei-c.org) as a possible background for its actual computer representation. Still there is quite a proportion of such scholars who would intuitively consider the TEI as not being fully appropriate for them, and sometimes even fearing that adopting the TEI may cause more trouble then benefit to their research project. This usually stems from a perception of the TEI as being both overly complex and at the same time under-empowered for dealing with the specificities of one's precise research.

In this context, the present paper is in no means intended to be a technical presentation of the intricacies of the TEI, but rather an informal overview of the general framework it offers, at a technical, but also an editorial and community point of view. We will thus try to see how the TEI may provide a valuable context for textual projects, identifying the first steps to go through to make an easy start with it, together with some practicalities that may just help any one to edit its first document within a quarter of an hour. By doing so, we expect that the reader who is already accustomed with the TEI may also see this presentation as the possible outline of what could be a first hand-on training session that you could have to make. Indeed, the questions that a newcomer has to face when starting with the TEI are usually those one has to address when welcoming a new contributor (student or scholar) to an existing TEI-based project.

Finally, I would want this paper to be an opportunity to demonstrate that the TEI exists because it has been put together not so much by techies, but by scholars themselves who, over the last twenty years, constantly tried to find the best compromise between scientific expectations and technical constraints.

## First step – you have a project in mind

Since you have read through this paper so far, it probably means that you had already some kind of an idea about a document or a corpus of documents that you wanted to have set in digital format, together with some kind of a purpose in mind as to its usage. It is actually important to consider, prior to the starting of any encoding project — even more to the choice of an encoding scheme — to which purpose the encoding activity is intended. The trade-of here is to provide a balance between the effort that one will put in encoding neatly many aspects of the source text (or the natively created digital text when applicable) and the actual benefit that can be driven out of this encoding from the point of view of legibility or processing. One of the main risk here is to tend towards an encoding over-kill, whereby a lot of effort is put in encoding precise phenomena in a text, which no other user than the encoder himself will ever have access to. In the remaining part of this section we will try to go through the main questions that one has to keep in mind when starting up an encoding project, namely, the objective of the project, what aspects are actually to be encoded, and probably most importantly what source material is available.

To start with, we can identify a few, yet probably non-exhaustive, dimensions along which one can answer the question: "what do you need encoding for?"

- Archival: when one wants to encode a text to make sure that this will remain legible within a far longer period of time than that of the encompassing project, it is necessary to consider that a) encoding standards and relevant documentation is potentially maintained for at least the same amount of time and b) that the level of encoding that has been put in the text is also relevant for a user beyond the period of the project;

- Scholarly work: this dimension introduces a trade-off which is to be clarified right from the onset of a project, namely which aspects of the encoding project will be specific to the contemplated research and which should in-turn actually be further kept for a wider distribution of the corpus;

- Dissemination: as soon as the corpus to be digitized is planned to be disseminated to a wider audience, one should make sure that the documentation of the corpus objects, both from a library point of view (meta-data, source identification, etc.) and a technical point of view (schema), is adequate for their autonomous processing by third-party users;

- Digital edition: when the corpus is planned to be used for providing an online or printed edition of the documents, one has to integrate in his work plan all features which will facilitate the presentation and handling (navigation) through the documents, or the actual presentational features that may not be inferable from a purely semantically oriented encoding of the text.

Once the purpose of the project is set, it is time to identify the features to be actually encoded, which boils down to identifying a coherent answer to the question: "What should I really (really) need to encode?" Such an answer can be devised along the following levels:

- Macrostructure of the documents: what is the general organisation of the documents in divisions? Do some of them have a specific role or nature (preface, letters, journal articles)? What internal structure (paragraphs, figures, poetic lines, bibliographical references, examples, mathematical formulas, etc.) should be actually encoded?

- Documentation: how precise should I link my digitized document to source related information, what will I trace specifically

in the course of my project (e.g. encoders' identification, revisions, used tools)?

- Surface annotation: in the light of the objectives assigned to the encoding project, what are the actual surface features that I should identify in my text (names, places, temporal expressions, external references, indices, etc.) and what will be the cost of keeping or disregarding the feature in terms of both budget and actual usefulness.

Whatever the actual aims and priorities set above, the final constraints on the digitisation project will come from the data itself, namely answering the question: "what is the available material?" Depending on the nature of the source, the actual priorities and resulting activities may vary quite importantly:

- Printed edition: this is probably one of the most usual cases in the field of digital humanities, especially for computational linguists[1]. The digitisation of such documents, whether manual or semi-automatic (OCR), usually leads to univocal content with shallow surface annotations (presentational features). The main issue is usually how far the actual encoding should transform surface features (e.g. italics) into more semantically oriented annotations (e.g. foreign expressions, titles, etc.). This is all in all an ideal case since one can really make an editorial choice as to the best match between editorial objectives and available material;

- Retro conversion of a digital source: Although this may be seen as an ideal situation, this usually implies more editorial constraints then the previous case. The range of possible formats (from typographically oriented LaTeX[2] to approximately conceived formats such as the NLM DTD[3]) brings in headaches on how to map the available features onto the format chosen for your project and deal with the trade-off between objectives and available information. One of the usual hidden difficulty is often also to retrieve enough information about the source (bibliographical information, copyright, etc.) to guaranty a good usability of the document;

---

[1] See for instance projects like the BNC (http://www.natcorp.ox.ac.uk/) or DTA (http://www.deutsches-textarchiv.de/).

[2] A document formatting system for the TeX typesetting program. See http://www.latex-project.org/

[3] A series of DTDs designed for the National Library of Medicine for the representation of journal article (see http://dtd.nlm.nih.gov/)

- Born digital: this is the usual situation for an editorial project such as an electronic journal or a reference document series (standards, patents, activity report) where the purpose of the text (mostly dissemination of information) is clearly known. As a result, the definition a finite set of features and corresponding practices is somehow simplified, with very little room for encoding overkill. Still, since the corpus of texts is a constantly evolving matter, there is a need for defining a workflow for constant updating of the underlying schema;

- Manuscript: Even if this case could have been set closer to that of printed material, it is indeed a highly peculiar and complex situation. Encoding manuscripts, whether they correspond to ancient sources or genetic documents, is mainly made more complex because a) the textual content is not necessarily easy to decipher, b) it may come with a variety of corrections and annotations and c) there are potentially many presentational features (seals, adornments, marginal layout) that one may want to keep because it impacts on the meaning of the content proper.

Even if most TEI based projects are closely anchored on digital humanities projects, it is becoming more and more common to see the TEI be used as the underlying framework for other types of documents. As an exemplification of the various aspects presented above, I would like to outline now the main characteristics of two non-scholarly projects for which the TEI is actually seen as the optimal framework for the management of their document structure.

### Case study 1: annual report of a research organisation

Context

INRIA, the French national research organisation in computer science[4], requires its research teams to produce an annual report of their research activities and production. This report is intended to serve two comple-mentary purposes. On the one hand, it is the basis for the assessment campaigns that take place every four years for each research team, and, on the other hand, it is published openly online as a vector for the wider dissemination of INRIA's activities[5].

---

[4] http://www.inria.fr.

Main characteristics of the documents

The document structure of such an annual report can be characterised along three main axes:

- It contains an extensive administrative section describing the members of the teams together with their periods of stay and actual affiliation[6]. This requires that persons, organisations and addresses are precisely described and encoded;
- The focus on research production implies that bibliographical references are precisely represented and classified (e.g. journal papers, conferences, workshops, reports) and related to both references in the text and possibly external material (data, software, online papers);
- The document structure is highly constrained in order to both provide a constant descriptive framework across research teams and ensure a predictable usage of the reports in further processing (online presentation, queries, automatic indicator extraction).

Editorial workflow

The usual trade-off for such a document type is to be able to provide coherent editorial guidelines, when, at the same time, the researchers are producing the content all by themselves and may thus introduce or even impose their own peculiarities. In particular, since the computer science community has a long-standing relationship with TeX, this rather presentational format has been chosen as the "natural" source format for authors'. The chapters, once proofread and finalized are then converted into an XML structure for archival and dissemination. Besides, some of the bibliographical information can — and in the long term, must — be

---

directly uploaded from the French national publication archive HAL[7] and be merged in the may text, prior or after its edition.

### Case study 2: back office format for a standardisation body

Context

ISO (International Organisation for Standardization[8]) is the major international standardisation body, federating the work of national bodies worldwide and covering basically all types of technical fields. ISO has published so far more then 17,000 standards, which at first were only distributed in paper format, and which progressively have been integrated within an electronic document workflow. For all standards ISO ensures that the content is the result of a consensus among participants in the corresponding technical committee, that it is properly referenced and distributed and also that it is regularly updated according to technological evolutions. ISO standards are mainly intended to be published in paginated form for reading, even if ISO explores databases as possible candidates for standardisation (e.g. language codes).

Main characteristics of the documents

ISO standards have a strict document organisation[9], which reflects the necessity for clearly identifying components such as *scope, terms and definitions, normative documents*, etc. They also come with a precise meta-data description stating the document title(s), the technical committee responsible for the preparation of the standard, the publication information (date, copyright, etc.). Besides, the variety of technical fields covered by ISO imposes that the content itself may contain many different types of objects such as graphics, formulas, technical drawings or specification code. In a way, ISO documentary base could be seen as the ideal playground for anyone who is interested in technical documentation.

Editorial workflow

ISO documents are usually edited by a small number of people (project editor possibly in relation with an editorial committee), which, being ex-perts in their own technical fields, do not have specific IT background beyond the basic usage of a word processor. As a result, most standard editing activities are operated in Microsoft Word with documents being disseminated as PDF's when ballots are taking place. At the final stage of the standard production phase the ISO central secretariat is manually converting the available document to produce an XML document to be integrated into the main ISO document management system.

---

[7] Hyper Articles en Ligne (http://hal.archives-ouvertes.fr/)

[8] http://www.iso.org/iso/home.htm

[9] See ISO directives part 2
  http://isotc.iso.org/livelink/livelink?func=ll&objId=4230456

Overview

The two projects briefly presented here are indeed typical cases where institutions are faced with the necessity to define a document format, which will be used for a large number of documents over a rather long period. This implies that the underlying document format, or schema, has to be reliably defined in such way that it is easy to be used, maintained and that it comes with a clear documentation. In the course of this paper we will see whether the TEI can offer such a framework and relate this analysis the actual history of both institutions in their endeavour to define such a format.

## Second step – you want to know more of the TEI

### *Theory*

At this stage in your thoughts, you have probably made even a tiny link with the existence of the TEI and want to come to terms with it. You have also had a few ideas or prejudices about it, which you would want to check against reality. As a matter of fact, you could recognise your thoughts in one or (more likely) all of the following statements:

- **TEI is based on XML**. This is probably a good feature and from what you have perceived of the global digital world, XML is now widely adopted by all communities, public or private, to represent any kind of information, whose structure or semantics is more important than its surface layout. Still, you may not know that indeed the TEI started before the XML era, but its founders had the idea to consider right from the beginning that SGML, the ancestor of XML, was at the time the best solution for controlling the organisation of a textual document. From its early technical activities, the TEI community managed to identify a lot of the features that were to become the core characteristics of XML;

- **It's too big.** This is usually the feeling that is conveyed to any-one just looking at the surface of the TEI guidelines and discov-ering that it offers more than 500 XML elements together with a thick documentation coming in 23 chapters, ranging from manuscript description to dictionary encoding. The feeling that even starting the simplest editorial project would require to go through all the corresponding prose naturally leads to an obvi-ous conclusion: you would better design your own XML DTD[10] or schema;

- **It's not enough.** As soon as you started to dive into the TEI guidelines and look for a specific issue, say, the affiliation of an author in a bibliographical representation, you might have im-mediately thought that you could not find exactly the kind of subtlety that is really needed for your project. At this stage you have come to an obvious conclusion: you should design your own XML structure.

We can actually make the preceding worries concrete by going through our two use cases and see how they positioned themselves. As one shall see, stabilizing a document editorial workflow is the result of a ripening process where one takes full benefits from past difficulties.

***Case study 1: annual report of a research organisation***

History

After a period during which INRIA annual reports were completely ed-ited as Tex documents, it became clear that the definition of a produc-tion line involving multiple output formats together with web accessibil-ity would require the use of a more content oriented format. XML very soon came up as the unavoidable choice, in particular in the context of INRIA being one of the three academic pillars of the W3C in the late 1990s. At that time, the importance of fully situating oneself within a standardised framework was not seen as a deep priority, in particular since the development of the underlying document scheme was itera-tively spread across several years. As a result, a self-made DTD was designed, which strongly inspired itself from the TEI framework while introducing specific construct that could be justified as follows:

- The report macrostructure was explicitly implemented by means of elements corresponding to all components needed for the evaluation of research activities: "identification", "presentation", "domaine", "logiciels", "resultats", "international", "diffusion", "biblio";
- A very precise structure was carved to deal with researchers' descriptions;
- Since most bibliographical data would be given by researchers as BibTex structures, a BibTex looking format was defined.

---

[10] Document Type Definition. The core language provided by the XML recommendation to express the syntax of an XML docu-ment and in turn to provide means to check the validity of a document against such a syntactic description.

Still the intermediate level tags (paragraphs, references) kept their TEI looking nature over the years while the format as a whole evolved continuously.

### Difficulties

The constant evolution of the document structure, together with the resulting lack of maintained documentation, created a situation where, first, tools had to be systematically updated to cope with the changes, and second, changes were made as small as possible (in the form of "patches") so that the whole editorial workflow would not break and prevent a timely production of the annual reports. The situation was made even worse when it was contemplated to refine the content to be able to produce precise research production indicators needed for institutional assessment.

### Perspectives

Given the context expressed so far and the difficulties that INRIA would face in changing its editorial workflow in haste, the best strategy that has been identified is to actually design a target document format, that is, an ideal document format (thus departing from the patch-syndrome) at which a corresponding evolution plan could aim. As a matter of fact it has been identified that the current document structure could be easily mapped onto a subset of the TEI guidelines and that by doing so, one could progressively switch older tools into TEI-aware components.

*Case study 2: back office format for a standardisation body*

History

Because of the need to provide precise access to standard document content, ISO introduced at a very early stage an SGML[11] back-office document structure. This allowed standards to be precisely checked at production time and potentially be fully exploited at a very fine-grained level of representation. The underlying document type definition was defined as a fully proprietary format closely sticking to ISO constraints. When XML came into play, the format was made compliant to the XML syntax without any major changes in its element set.

Difficulties

One constant feeling in ISO is that there has always been a strong discrepancy between the editing process of standards within ISO committees and the final production line. In particular, nothing facilitates the conversion of committee-produced documents into the ISO XML structure. Besides, just like for INRIA, the proprietary nature of the ISO format induced difficulties both of documentation maintenance and tool update when new features would come into play (for instance when new technical domains would be tackled within ISO).

Perspectives

Since December 2007, ISO central secretariat has decided to design a new document workflow that would both be based on a standardised structure and provide a smooth transition between technical committees and the final production of standards. However, the need of keeping a content oriented structure rather then a presentational one made them consider the TEI as an ideal framework to this end[12]. The main arguments that ISO put forward were the following ones:

- The completeness of the TEI element sets that covers most of the features needed for standard editing and production;

---

[11] ISO standard 8879

[12] In particular in comparison to more layout oriented standards such as ISO/IEC 26300 (Open Document Format for Office Applications (OpenDocument)), or the ongoing proposal ISO/IEC DIS 29500, Office Open XML file formats.

- The modular architecture together with the customization facilities;
- The precise, and multilingual documentation;
- The very active community ensuring a very good reactivity to technical evolutions.

## Getting started

Let us now see why the TEI does allow one to prevent from having the abovementioned difficulties and how it is possible to have a quick grasp on the technical content.

First, it is necessary to understand what the TEI actually brings.

From a general point of view, it gives you the means to define the logics of your own text. That is, it gives you some *guidelines* for identifying what kind of structural object (e.g. division in a text, paragraph, sentences) or specific phenomena (identification of names and places, precise bibliographical references, metrical structure of a poem) you actual want to identify and markup. It also gives a strong background to document your work along various dimensions like the identification of the source document (where you found your source manuscript), the actual participants in the dialogue you are transcribing, or the tracing of the various versions of your encoding work.

But more precisely, the TEI offers a comprehensive background for actually making your own choices concrete from a technical point of view. If you happen to discover that you want to integrate dictionary entries in your otherwise prose document you can actually add to your document model the module named "Dictionaries"[13].

In order to actually start working on the edition of a TEI compliant document, one actually needs to have the following element ready at hand:

- A "good" XML editor to control the edition of the XML file being created according to a TEI compliant schema. One would without any risk recommend <oXygen/>[14] to this purpose which provides all functionalities (RelaxNG validation, full

---

[13] http://www.tei-c.org/release/doc/tei-p5-doc/en/html/DI.html
[14] http://www.oxygenxml.com/

> XSLT transformation support) that one needs for a comfortable experiment;

- An access to Roma under http://www.tei-c.org/Roma/, in order to define the TEI variant one want to use (see below) and generate the corresponding schema;

- An access to the TEI documentation under http://www.tei-c.org/release/doc/tei-p5-doc/en/html/index.html to consult both the general prose about the various TEI modules or the precise descriptions of each element of the guidelines.

Once this environment is available, the actual work can start by choosing the TEI customisation most suitable to your needs.

## Two basic scenarios

### TEI absolutely bare

The *TEI absolutely bare* schema would be the one to recommend to start with for discovering the TEI from scratch and with the minimal set of possible elements. It is available from Roma[15], from the list of predefined customisations under "Create customization from template" (click 'start'). Once there, you can download ("schema" tab) a RelaxNG schema and start editing an XML document on your favourite editor accordingly. This is indeed enough to generate a first document such as the one shown in figure 1 below. This documents demonstrates some of the properties of the TEI:

- It groups together both the document content proper (in the <text> element), but also the metadata attached to it (<header> element), so that a TEI document is a completely autonomous digital object that can be archived, transferred or manipulated independently of any extra third party information[16];

- The header itself, contrary to some other digital metadata initiatives such as the Dublin Core initiative[17], comes as a highly structured component allowing to clearly group together, like in this example, information pertaining to document identification (<titleStmt>), dissemination (<publicationStmt>) or origin (<sourceDesc>);

---

[15] http://www.tei-c.org/Roma/

[16] The elicitation of the relation between the document and the actual schema used to validate it may make this assertion only partially true. Discussing this (important) point goes beyond the scope of this introductory paper.

[17] ISO/IEC DIS 29500

- The macro-structure of a document is based on a quite usual organisation of textual content (the <front>, <body>, <back> structure) and generic structural objects for the representation of divisions in a hierarchical manner (<div>, with <head>);
- Textual content is in turn organised into semantic units such as paragraphs (<p>) or lists (<list>) that bear no presentational prejudice.

```xml
<?xml version="1.0" encoding="UTF-8"?>
<?oxygen
RNGSchema="file:/Users/romary/Downloads/tei_bare.rnc
" type="compact"?>
<TEI xmlns="http://www.tei-c.org/ns/1.0">
    <teiHeader>
        <fileDesc>
            <titleStmt>
                <title>My first TEI document</title>
                <author>Laurent Romary</author>
            </titleStmt>
            <publicationStmt>
                <p>Distributed under CC-BY</p>
            </publicationStmt>
            <sourceDesc>
                <p>Born digital document</p>
            </sourceDesc>
        </fileDesc>
    </teiHeader>
    <text>
        <front>...</front>
        <body>
          <div>
              <head>A division with a title</head>
              <p>Demonstrated here is that:</p>
              <list>
                  <label>Main argument</label>
                  <item>The TEI is very simple</item>
                  <label>Even better argument</label>
                  <item>The TEI is elegant</item>
              </list>
          </div>
        </body>
        <back>...</back>
    </text>
</TEI>
```

Figure 1: Minimal TEI document.

### *A real TEI application*

To illustrate the kind of constructs that allow fine-grained representations of complex entities in a TEI text, let us consider the case of sequences of references to persons, together with some characteristics attached to them. This is typically needed when encoding the various interlocutors in a dialogue to be transcribed, or, in the use case we have been dealing with so far, to record the participants to an organisational unit such as a research group at INRIA. To this purpose, the TEI offers the <listPerson> element, which groups together a sequence of <person>'s, together with additional organisations or relations that these persons are part of or involved in. In the case of a research report we can make use of a quite elaborate instance of this construct by providing additional information related to the affiliation of the participants, as well as their academic status of location.

To achieve this we can select on Roma the "build schema" option and add (tab: modules) the Names and dates module[18]. Once done, the RelaxNG schema resulting from this customisation allows one to describe person list such as the one below.

```
<listPerson type="staff">
    <person>
        <persName>
            <forename>Malte</forename>
            <surname>Rehbein</surname>
        </persName>
        <affiliation>
            <orgName type="university">National Univer-
sity of Ireland, Galway</orgName>
            <orgName   type="department">Moore    Insti-
tute</orgName>
            <state type="grant">
                <desc>Marie Curie Research Fellow</desc>
            </state>
        </affiliation>
    </person>
    <person>...</person>
    <person>...</person>
</listPerson>
```

---

[18] As documented in http://www.tei-c.org/release/doc/tei-p5-doc/en/html/ND.html

## The hidden faces of the TEI

While reading this paper, you have probably now reached a stage where you have already managed to come across some useful information to go forward in your project, downloaded an XML editor and maybe even opened your first TEI document after a short trip to Roma. However, you are still facing some kind of difficulties in understanding how you are going to match the full documentation available to you on the TEI web site and your own constraints within your project. You may even have come across some specific encoding situations where it seems that your own thoughts are already going beyond what is available in the TEI guidelines. This is where it is necessary for you to get acquainted with what is probably the most useful tool within the TEI framework, namely its user community.

The TEI guidelines should indeed only be considered in the light of the group of people who in the last twenty years have put in their scientific and technical expertise to create the sound platform that we have at present. This community is now as active as ever since the foundation of the TEI consortium, which is now the host of all editorial and technological developments. The TEI consortium relies on the membership of organisations having whatever kind of interest in the TEI, and is organised on a board[19] for organisational matters and a council[20] for technical ones.

The communication between the consortium and the TEI community is channelled through several places:

- The official TEI web site[21], which contains most of the "stable" information about the TEI activities and the TEI guidelines;

- The TEI wiki[22], which acts as a working place where TEI related information (e.g. project descriptions) is presented and SIGs (Special Interest Groups) find their home base;

- TEI@sourceforge[23], where the source files (guidelines, stylesheets, software) are being maintained by both the council and the community at large;

---

[19] http://www.tei-c.org/About/board.xml

[20] http://www.tei-c.org/Activities/Council/

[21] http://www.tei-c.org

[22] http://www.tei-c.org/wiki/index.php/Main_Page

[23] http://sourceforge.net/projects/tei/

- Roma[24], the online environment for managing and accessing TEI compliant schema;
- Last but not least, the TEI mailing list[25], which is probably the core exchange forum of the TEI community.

It is important to notice that the dynamicity of the mailing list, where both newcomers and techies can actually find their benefit, comes from the actual variety of origins of the people involved in TEI related projects.

To illustrate this, we can go through a discussion thread that took place on the TEI list in March 2008[26] on the issue of describing the internal structure of names. I initiated the thread by asking the following question.

> "While defining encoding guidelines for an institutional bibliographical list, I came across the issue of recording both the first name of an author ('Carlos') and the corresponding abbreviated form ('C.'). I am considering several possibilities to deal with this in a systematic way and would like some advice/hints/comments about this.
>
> The encoding context is:

```
<author>
  <persName>
        <forename>Carlos</forename>
        <surname>Areces</surname>
  </persName>
</author>
```

> The possibilities I see are:
>
> a. tell the institution to forget about this information and compute it on the fly
>
> b. use a typed forename

```
<persName>
  <forename>Carlos</forename>
  <forename type="initial">C.</forename>
  <surname>Areces</surname>
```

---

[24] http://www.tei-c.org/Roma/

[25] listserv.brown.edu/archives/tei-l.html

[26] Thread starting under: http://listserv.brown.edu/archives/cgi-bin/wa?A2=ind0803&L=TEI-L&P=R6440

```
</persName>
```

c. provides a normalized (for the institution) representation by taking up @norm from att.lexicographic

```
<persName>
  <forename norm="C.">Carlos</forename>
  <surname>Areces</surname>
</persName>
```

d. introducing a new element (maybe a synonym for b.)

```
<persName>
  <forename>Carlos</forename>
  <initial>C.</initial>
  <surname>Areces</surname>
</persName>
```

This quite specific question generated a whole range of answers, which I have slightly edited here and which actually exemplify the variety of views that one can get for such an issue.

The most pragmatic and quickest answer came from Sebastian Rahtz, the main hand behind the TEI technology:

> "a. tell the institution to forget about this information and compute it on the fly
>
> yes please.
>
> ```
> <forename rend="initial">Carlos</forename>"
> ```

Peter Boot, whose work on emblems[27] has given him experience in going beyond the simplicity of primary questions elaborates:

> Any solution will have to distinguish between this situation (the initial is an alternative to the full forename) and the situation where the initial corresponds to the middle name, as in
>
> ```
>   <persName>
>     <forename>George</forename>
>     <forename type="initial">W.</forename>
>     <surname>Bush</surname>
>   </persName>
> ```
>
> What comes to mind is <choice>...
>
> ```
>   <persName>
> ```

---

```
    <choice>
       <forename>Carlos</forename>
       <forename type="initial">C.</forename>
    </choice>
    <surname>Areces</surname>
  </persName>
```

Alexey Lavrentev, whose works on the transcription of Slavic folklore, has brought him to detect anomalous encoding provides the right encoding for Peter's initial intuition:

> The content model of <choice> is actually quite restrictive and does not allow <forename>.

> A solution might be to use <abbr> and <expan>

```
<persName>
    <forename>
        <choice>
            <expan>Carlos</expan>
            <abbr>C.</abbr>
        </choice>
    </forename>
    <surname>Areces</surname>
</persName>
```

Still, probably pondering on the complexity of the achieved solution, Alexey Lavrentev continues with a further contribution:

> "I wonder whether it is necessary to specify the initial when the full form is provided: the operation is mechanical and can easily be done at the moment of visualization using XSLT or some script with regular expressions."

Arianna Ciula her precise of the knowledge gained both at King's college and in the TEI council could not resist bringing a simplified feature:

> What about using @full? [quoting the guidelines] "The @full attribute may be used to indicate whether a name is an abbreviation, initials, or given in full"

```
<persName>
  <forename>Carlos</forename>
  <forename full="initial">C.</forename>
  <surname>Areces</surname>
</persName>
```

Immediately refuted by Sebastian Rahtz, though, for sake of ambiguous semantics:

Useful, but the encoding suggests (at least to me) that this author is named Carlos C. Areces.

We were obviously missing here the scholarly view that only Matthew Driscoll, a long-standing specialist of Scandinavian manuscript could bring:

"This could be slightly problematic in the case of languages that treat initial consonant clusters as a unit, for example most of the Germanic languages (though not English, which stopped being a proper Germanic language long ago), where the 'initial' of a name like Christian would be Chr., Thomas Th. and so on."

Do-able through @rend, I'm sure, just slightly problematic.

Syd Bauman, former co-editor of the TEI guidelines, sees the problem through its various options:

I am with Sebastian or Alexey LAVRENTEV or personography here. Either don't record this information, use <choice>, or rely on your personography.

```
<persName>
    <forename>
        <choice>
            <expan>Carlos</expan>
            <abbr>C.</abbr>
        </choice>
    </forename>
    <surname>Areces</surname>
</persName>
```

seems perfectly adequate. You could, of course, record this once in the personography and just use

```
<persName ref="persons.xml#careces.tsj"/>
```

in your TEI bibliography. Then you can go to town with the details of the <persName> inside the <person xml:id="careces.tsj"> element only once.

Bringing in his experience of TEI based editing[28], Martin Holmes goes beyond the technical answers to provide a pragmatic background:

"It's not always possible to render initials automatically from full names, without extra info. Many people "spell" their names in

---

[28] http://hcmc.uvic.ca/blogs/index.php?blog=30

> idiosyncratic ways, with odd uses of case -- for instance, 'k.d. lang'
> http://www.kdlang.com/home.php where there's no space be-
> tween the initials, and they're lower case. I think there needs to be
> a mechanism for specifying initials in whatever form is preferred,
> rather than trying to guess at them programmatically."

With the final, though elliptical, word to Lou Burnard:

```
"<forename type="initials">e e</forename>
<surname>cummings</surname>[29]
```

> springs also to mind, archy[30]"

I ended up finding both a good and simple answer for my application
(using the @rend attribute, to indicate how I would want to see the re-
sult presented) and having a clear picture on the various issues that I
should take into account for other projects where the use case would be
slightly different. This overall spectrum from simple mechanisms to
scholarly complexity is a exemplary of the way the TEI works.

## Conclusion

The TEI has proven over the years to be one of the very few communi-
ties where both beginners and advanced users can actually exchange very
precise technical information. This is essentially due to the fact that this
community shares a common interest and culture about what an elec-
tronic text can be and behind any technical request always lies a query
about the actual nature of a textual feature.

One further characteristic of the TEI resides also in its capacity for
change and evolution while maintaining its core underlying principles.
The wide coverage of the guidelines results indeed in users constantly
finding constructs which do not fit actual usage and which have to be
better taken care of. Still, such changes are always dealt with as sources
for generalisation, so that a solution found here may also be seen as an
improvement elsewhere in the guidelines.

The main challenge for the TEI is to be able to keep this coherence
while allowing specific projects or communities to derive fine tuned cus-
tomization for their own usage. In this respect, the ISO back office pro-
ject is exemplary since it results in a very strict organisation of possible
TEI element, but at the same time an excellent example of the kind of applications that many other settings may want to achieve. In this context, the TEI consortium has the duty to record such application profile and offer them as a second layer of technical information, more application oriented, which is widely disseminated to the community as a whole.

---

[29] http://en.wikipedia.org/wiki/E._E._Cummings

[30] http://www.donmarquis.com/archy/

## References

No specific reference is given but the bibliography maintained on the TEI website under:
http://www.tei-c.org/Activities/SIG/Education/tei_bibliography.xml

## Coordinates

Laurent Romary
INRIA
Rheinsberger Strasse 79
10119 Berlin
laurent.romary@loria.fr